\documentclass[a4paper]{spie}  

\usepackage{amsmath,amsfonts,amssymb}
\usepackage{graphicx}
\usepackage[colorlinks=true, allcolors=blue]{hyperref}

\title{Future exoplanet direct imaging instruments:
Simulating spatial light modulator-based pixelated focal-plane coronagraphy}

\author[a]{L. Lin}
\author[a]{A. Potier}
\author[a]{R. Tandon}
\author[a]{J. Kühn}

\affil[a]{Division of Space and Planetary Sciences, University of Bern, Silderstrasse 5, 3012 Bern, Switzerland}

\authorinfo{Further author information: (Send correspondence to L.Lin)\\L.Lin: E-mail: liurong.lin@unibe.ch, Telephone: +41 31 684 85 76}

\begin{document} 
\maketitle

\begin{abstract}

The programmable Liquid-crystal Active Coronagraphic Imager for the DAG Telescope (PLACID) instrument will be installed on the Turkish 4-m Telescope by the fall of 2024 and is expected to be on-sky by the end of the year. PLACID will be the first “active stellar coronagraph instrument”, equipped with a customized spatial light modulator (SLM), which performs as a dynamically programmable focal-plane phase mask (FPM) from H- to Ks- band. A Python-based numerical simulator of SLM-based focal-plane phase coronagraph is developed to investigate the effects of discrete pixelated FPM patterns in place of classical phase masks. The simulator currently explores the impacts of two design choices, spatial sampling in the coronagraphic focal-plane (number of SLM pixels per $\lambda$/D) and phase resolution (SLM greylevel steps). The preliminary results of the monochromatic simulations show that in ideal conditions (no wavefront errors) it is sufficient to use FPMs with spatial sampling of 10 SLM pixel per $\lambda$/D and phase resolution of 8 bits. The tool is expected to enable detailed simulations of PLACID or similar SLM-based instruments, and to help with real-time operations (optimal choice of FPM for given observing conditions) and interpretation of real data. Additionally, the tool is designed to integrate and simulate advanced operation modes, in particular focal-plane phase diversity for coherent differential imaging (CDI) of exoplanets.

\end{abstract}


\section{INTRODUCTION}
\label{sec:intro}
The past few decades have witnessed over 5000 confirmed detections of exoplanets and candidates through indirect methods. However only a few dozens of them have been directly imaged with High contrast imaging techniques (HCI) \cite{Lagrange_2010} due to technical limit of achieving high contrast at small angular separation, especially with ground-based instruments where post-adaptive optics wavefront residuals are highly detrimental. In high contrast direct imaging, the unwanted starlight suppression is usually achieved by using an internal coronagraph instrument. A coronagraph helps to reveal the planet signal by either blocking or diffracting the starlight in the first focal plane by changing the amplitude or phase of the incoming light using a focal-plane phase mask (FPM). Commonly used focal-plane phase masks are vortex phase masks \cite{Vortex}, four-quadrant phase masks (FQPM) \cite{FQPM} and Roddier $\&$ Roddier phase masks \cite{Roddier}. With a different charge number, the properties of vortex phase masks change. As the charge number increases, the vortex phase mask is less sensitive to the wavefront errors, especially tip/tilt jitter, while the lower charge number improves the throughput on off-axis companions very close to the host star.  \cite{Ruane_2018}.

As different FPMs exhibit different advantages and drawbacks in the coronagraphs, optimally adapting different FPMs based on different observing conditions and targets becomes significantly important. A liquid-crystal on-silicon spatial light modulator (LCOS SLM) can change the refractive index of the linearly-polarized light, which subsequently changes the phase shift values of the reflecting light per pixel\cite{SLM}.  This property enables the replacement of the classical fixed FPMs with a LCOS SLM as a remotely programmable FPM \cite{Kuhn_2016}. The programmable Liquid-crystal Active Coronagraphic Image for the DAG Telescope (PLACID) for the 4-m Turkish DAG telescope will be the first “active stellar coronagraph”, equipped with a customized SLM, which will perform as a dynamically programmable FPM from H- to Ks- band \cite{Kuhn_2022}. This numerical simulation work is to mainly explore the effects of discrete pixelated phase masks. The simulation details are given in Section \ref{sec:method}, and the results with different parameters of the FPMs are shown in Section \ref{sec:result}.

\section{METHODS}
\label{sec:method}
To explore the impacts of various design parameters to generate a pixelated focal plane phase mask with an SLM, a python-based tool box is built. For the simulation of the coronagraph instrument, 2-D Fast Frourier transfer (FFT) is used for the light propagation. The padding multiplier is set to be 100 to increase the quality of the FFT results. The following assumptions are made to simplify the currently presented initial simulation. Firstly, only monochromatic light is implemented in the simulation. When simulating the FPMs, perfect 0-2$\pi$ linear phase mapping is assumed for n-bit grey levels. Furthermore, no wavefront errors for the in-coming light aberration are considered. Lastly, the SLM panel 2nd order effects such as fill factor, phase jitter, crosstalk, ghost reflection etc. are neglected in the simulation.

\begin{figure} [h!]
   \begin{center}
   \begin{tabular}{c} 
   \includegraphics[height=5.5cm]{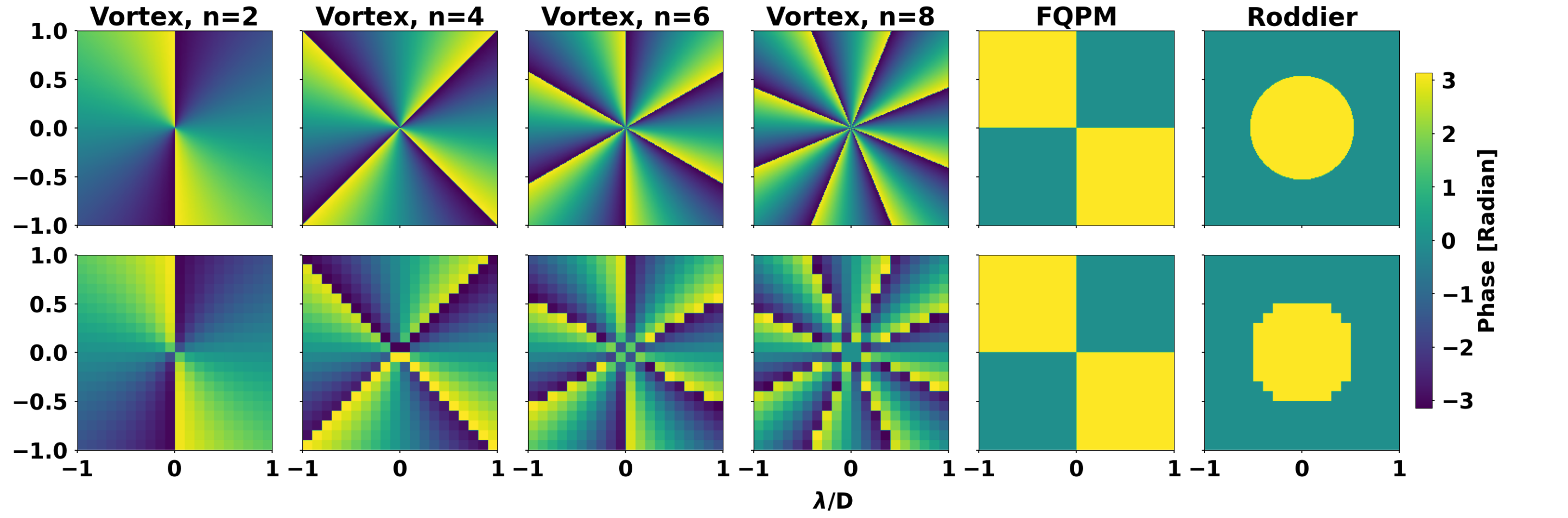}
   \end{tabular}
   \end{center}
   \caption{Examples of most commonly used FPMs with two different spatial samplings in the focal plane. The \textbf{Top} row FPMs are displayed with 100 pixels per $\mathrm{\lambda}$/D, while the \textbf{Bottom} row FPMs are generated with 10 pixels per $\mathrm{\lambda}$/D.}
   { \label{fig:ss} 
}
   \end{figure}

One of the limitations of using SLMs is the number of spatial sampling in the coronagraphic focal-plane, defined as the number of SLM pixels per telescope resolution element ($\lambda$/D). Currently, most commercially available SLM panel contain 1-2 million pixels. The configuration for SLM phase masks is  10 pixels per $\mathrm{\lambda}$/D. Limited by the available computer memory, the maximum number of spatial sampling allowed in the simulation is 100 pixels per $\mathrm{\lambda}$/D. This is used as an ideal case to compare with the performance of actual SLM phase masks with coarser sampling. Figure \ref{fig:ss} shows the examples of most commonly used FPMs vortex phase masks (charge = 2, 4, 6, 8), FQPM and Roddier $\&$ Roddier with the two different spatial samplings mentioned above.

\begin{figure} [h!]
   \begin{center}
   \begin{tabular}{c} 
   \includegraphics[height=7cm]{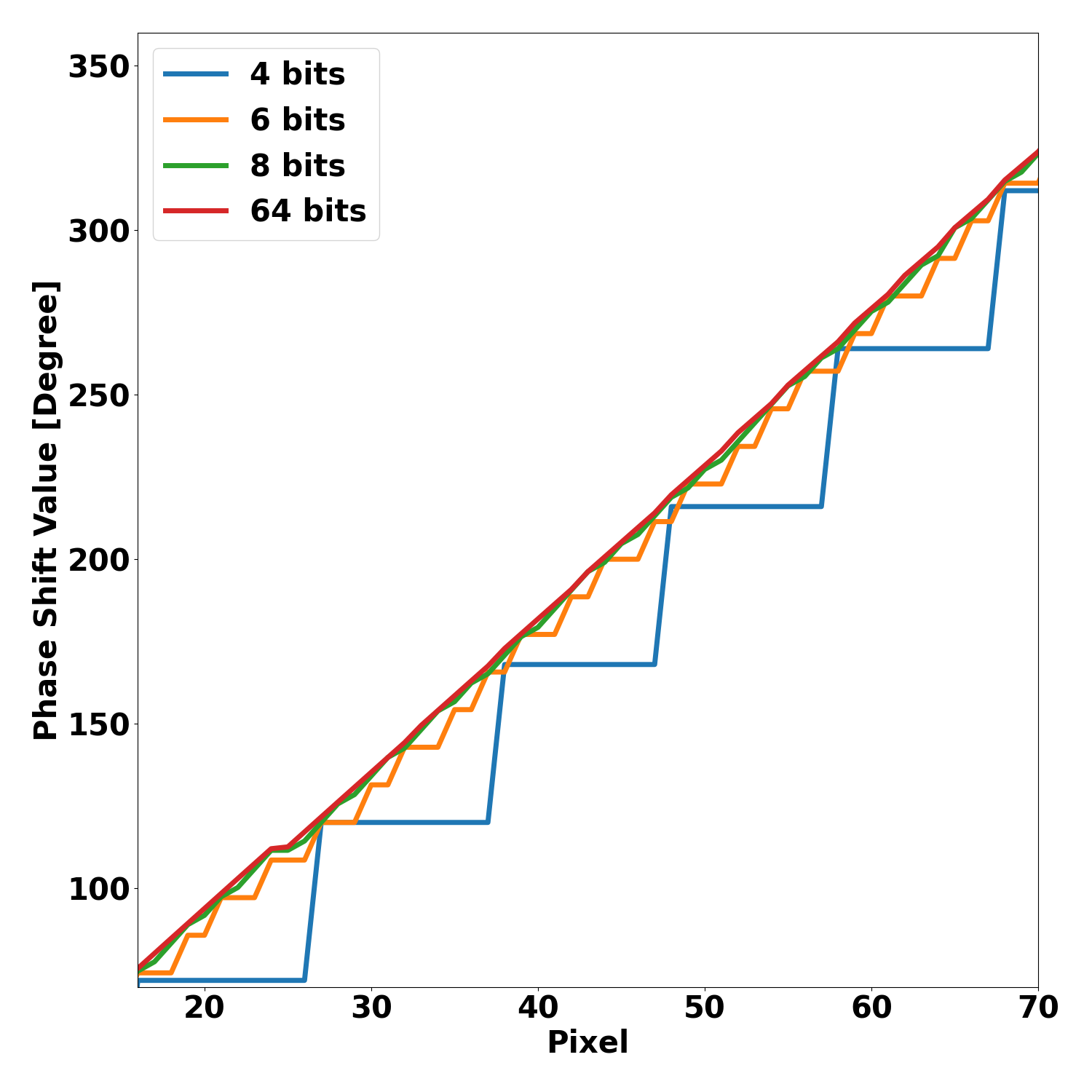}
   \end{tabular}
   \end{center}
   \caption{Impact of limited phase resolution for generating e.g. a phase ramp. Most commercially-available SLM panels(including the one in PLACID) have 8-bit resolution(10-bit in rare instances)}
   { \label{fig:gs} 
}
   \end{figure}

Another limitation is the limited phase resolution for generating the phase ramp. For the general numerical simulation in python, 64-bit memory addresses are used. However, the most commercially-available SLM panels are working with only 8-bit phase resolution (256 grey levels). Figure \ref{fig:gs} shows an example of the phase ramp from 70 to 360 degrees with different grey scale levels.  Most commercially SLM panels are equipped with 8 bits.    

In order to evaluate the performance of the phase masks with different spatial samplings and phase resolutions, the post-coronagraphic science images with different FPMs are normalized to the peak intensity of the post-coronagraphic science image without FPMs installed (non-coronagraphic PSF), and this translates into a normzalized post-coronagraphic PSF, where each pixel has a raw contrast value.

When considering more realistic telescope apertures with a central obstruction (secondary mirror), the sizes of the obstruction lyot stops can be another factor that impacts the final results. The DAG telescope is with a primary mirror of D$_{p}$ = 3.94 m and secondary mirror of D$_{s}$ = 0.975 m (D$_{s}$/D$_{p}$ $\approx$ 0.25) \cite{Kuhn_2022}. Lyot stops are usually oversized at the secondary mirror region to help with contrast and to migrate pupil registration errors. In this simulation, the following lyot stop central obstruction oversizing factors are implemented,  D$_{s'}$/D$_{s}$ = 1, 1.2 and 1.3. The bottom row of Figure \ref{fig:lyot_plane}
shows the visualization of different sizes of lyot stops in the lyot planes for five types of FPMs.

\section{RESULTS}
\label{sec:result}
\begin{figure} [h!]
   \begin{center}
   \begin{tabular}{c} 
   \includegraphics[height=5cm]{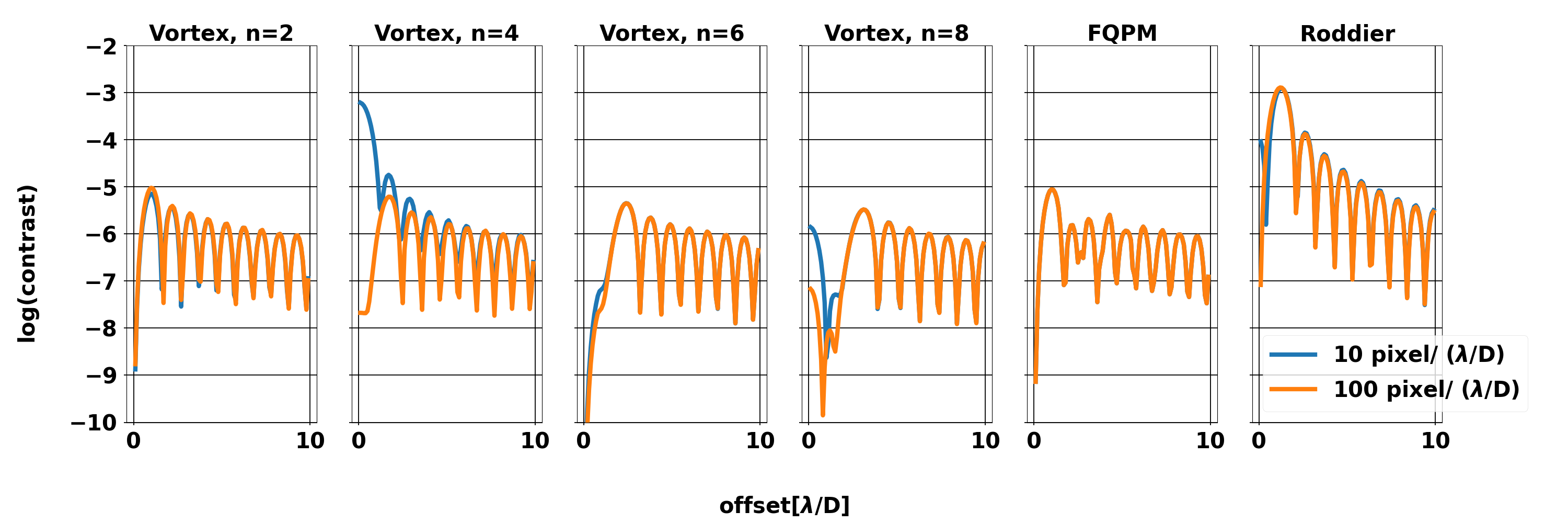}
   \end{tabular}
   \end{center}
   \caption{Post-coronagraphic focal-plane contrast for 6 different FPMs with 10 and 100 pixels per $\lambda$/D each of Figure \ref{fig:ss}}
   { \label{fig:contrast_ss} 
}
   \end{figure}

To better visualize and compare the post-coronagraphic science images with different FPM parameters, azimuthally averaged contrasts are computed for the normalized post-coronagraphic science PSFs mentioned in Section \ref{sec:method}. The results are plotted in Figure  \ref{fig:contrast_ss}. For most of the phase masks that are presented here, a sampling of 10 SLM pixels per $\mathrm{\lambda}$/D seems sufficient in performance. The contrasts of vortex phase masks with charge 4 and 8 are most affected by the lower spatial sampling. This can be attributed to the missing four quadrant patterns in the very center region of vortex phase masks with charge 4 and 8, which can be found in vortex phase masks with charge 2 and 6. 

\begin{figure} [h!]
   \begin{center}
   \begin{tabular}{c} 
   \includegraphics[height=5cm]{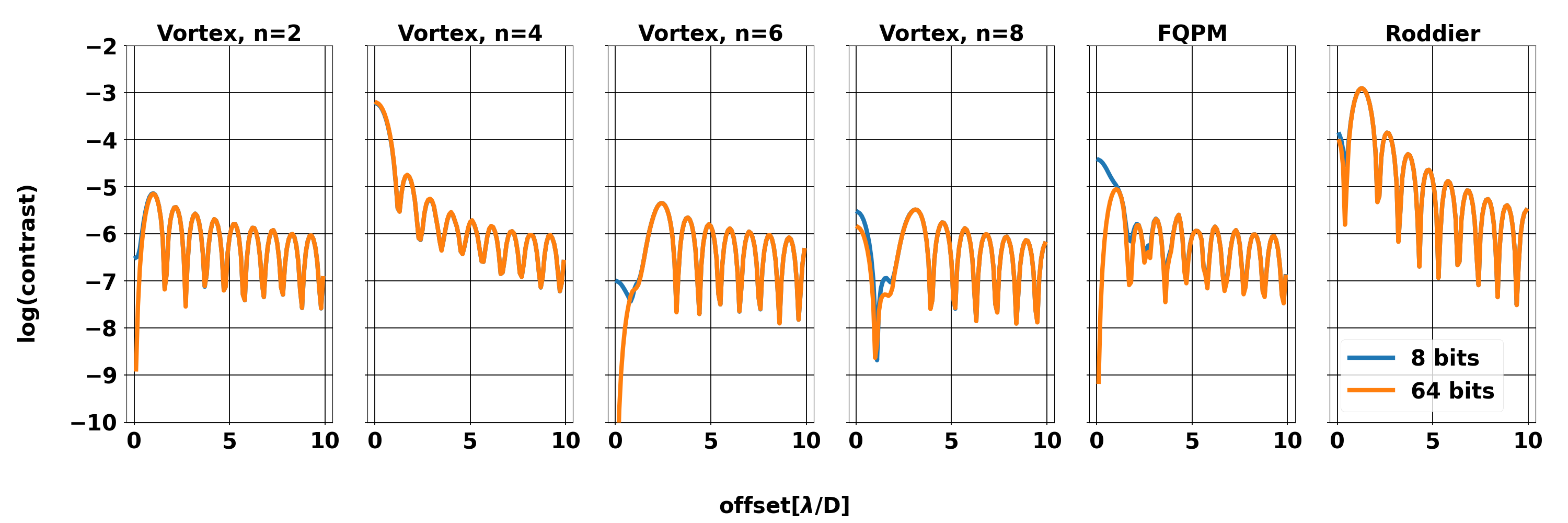}
   \end{tabular}
   \end{center}
   \caption{Post-coronagraphic focal-plane contrast curves for the 6 FPMs of Figure \ref{fig:ss} bottom row (10 pixels per $\lambda$/D) for 8 bits and 64 bits (Python floating number precision).}
   { \label{fig:contrast_gs} 
}
   \end{figure}

Figure \ref{fig:contrast_gs} shows the post-coronagraphic focal-plane contrast curves of the 6 FPMs with spatial sampling of 10 SLM pixels per $\mathrm{\lambda}$/D with 8 bits and 64 bits. There is no significant contrast penalty showing for using phase grey scale of 8 bits. FQPM is most affected by the greyscale level, because there are only two values in the phase mask and a slight variation of the phase value due to lower phase resolution possibly have a strong impact. 

\begin{figure} [h!]
   \begin{center}
   \begin{tabular}{c} 
   \includegraphics[height=10cm]{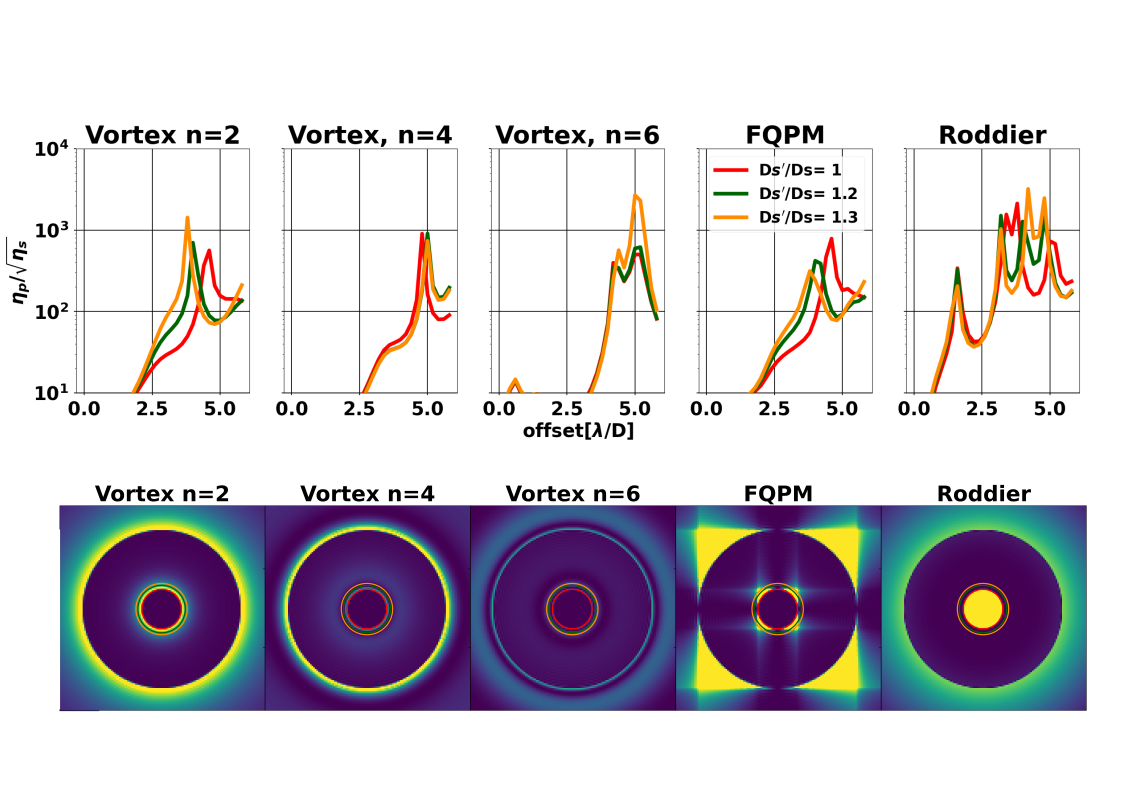}
   \end{tabular}
   \end{center}
   \caption{\textbf{Top}: $\eta_{p}$/$\sqrt{\eta_{s}}$ with different lyot central obstruction oversizing factors Ds'/Ds for the DAG telescope pupil, with $\eta_{p}$ being the throughput on the planet. and $\eta_{s}$ the one of the central star. \textbf{Bottom}: Post coronagraphic intensity distribution in the lyot pupil plane (before the lyot stop), with the coloured circles depicting the lyot stop secondary masks for the various case. add color illustration here. }
   { \label{fig:lyot_plane} 
}
   \end{figure}

To evaluate the performance of the coronagraph with different FPMs after adding the secondary mirror, the throughput of light after the focal plane phase masks should also be considered. In particular, the optimal oversizing factor for the secondary mask in the lyot plane is considered, trading contrast for throughput. The ratio of the throughput of the planet and the square root of the throughput of the star, $\alpha = \eta_{p}/\sqrt{\eta_{s}}$ is considered to be proportional to the signal to noise ratio of the companion \cite{Ruane_2018}. The top row of Figure \ref{fig:lyot_plane} shows $\alpha$ from 0 to 6 $\lambda$/D for 5 different FPMs with different lyot stop central obstruction oversizing factors, D$_{s'}$/D$_{s}$ as mentioned in Section \ref{sec:method}. The performances of vortex phase mask with charge 2 and FQPM are most affected under different sizes of lyot stops, because of the non-negligible amount of residual stellar light in the vicinity of the central obstruction. The undersizing of lyot stop central obstruction does not show too much impact for the Roddier $\&$ Roddier phase mask, because the lyot plane intensity images show that all the leakage light for Roddier $\&$ Roddier gathers inside the region of the secondary mirror.

\section{CONCLUSIONS and OUTLOOK}
\label{sec:con}

From the simulation, a conclusion can be drawn that choosing an SLM with spatial sampling of 10 pixels per $\lambda$/D and phase resolution of 8 bits (chosen parameters of SLM) to replace a classical FPMs does not degrade the contrast performance of the SLM at least under the current assumptions of monochromatic light and a perfect wavefront. Each FPM coronagraph has a different optimal lyot stop, thus the lyot stop sizing should be adapted to different FPMs, although it is foreseen that the introduction of wavefront error terms may change the optimal trade-off parameters.

One of the future developments of the simulator will be to introduce the broadband light (typically 20 $\%$ bandwidth) conditions, which will be an important step in the direction of a realistic simulator, notably in the context of the PLACID instrument. The chromaticity of the SLM-generated scalar FPMs is expected to worsen the final contrast to a significant degree when broadband images are considered. Innovative FPMs such as radial phase mask dimple \cite{dual_zone} and wrapped vortex phase masks \cite{wrap_vortex} have been proposed to improve the performance of traditional vortex phase masks under broadband condition. Other new types of FPMs are planned to be developed during the simulation to achieve a optimal broadband imaging results.  When PLACID is on sky, expectedly by the end of 2024, a realistic post-adaptive optics wavefront error residual budget will be available, as well as other instrumental parameters (pupil registration errors, tip/tilt jitter etc.)will be obtained. Adding the impacts of wavefront error, SLM calibration errors, pixel level phase jitters and pixel crosstalk etc. into the simulation and comparing the simulation contrast to the on-sky contrasts can provide significant information for the interpretation of the real time data. The SLM is expected to be able to be manipulated at specific modulation frequency. Therefore, zonal temporal phase-shifting for time-domain Cohererent Differential imaging (CDI) is one of the advanced operation modes that is going to be integrated into the simulations, notably to evaluate the optimal modulation scheme spatially (zone size and shape) and temporally (speed and number of phase steps). Indeed, time domain CDI has the potential to dynamically disentangle coherent residual starlight from incoherent off-axis astronomical sources of interest (point-source companions, disks).

\section{ACKNOWLEDGEMENTS}
\label{sec:ac}
The RACE-GO project has received funding from the Swiss State Secretariat for Education, Research and Innovation 
(SERI), under the ERC replacement scheme following the discontinued participation of Switzerland to Horizon Europe. 
Part of this work has been carried out within the framework of the National Centre of Competence in Research PlanetS 
supported by the Swiss National Science Foundation under grants 51NF40 182901 and 51NF40 205606. 

\newpage
\bibliography{report} 

\begin{thebibliography}{10}

\bibitem{Lagrange_2010}
{Lagrange}, A.~M., {Bonnefoy}, M., {Chauvin}, G., {Apai}, D., {Ehrenreich}, D.,
  {Boccaletti}, A., {Gratadour}, D., {Rouan}, D., {Mouillet}, D., {Lacour}, S.,
  and {Kasper}, M., ``{A Giant Planet Imaged in the Disk of the Young Star
  {\ensuremath{\beta}} Pictoris},'' {\em Science}~{\bf 329},  57 (July 2010).

\bibitem{Vortex}
{Mawet}, D., {Serabyn}, E., {Liewer}, K., {Burruss}, R., {Hickey}, J., and
  {Shemo}, D., ``{The Vector Vortex Coronagraph: Laboratory Results and First
  Light at Palomar Observatory},'' {\em \apj}~{\bf 709},  53--57 (Jan. 2010).

\bibitem{FQPM}
{Rouan}, D., {Riaud}, P., {Boccaletti}, A., {Cl{\'e}net}, Y., and {Labeyrie},
  A., ``{The Four-Quadrant Phase-Mask Coronagraph. I. Principle},'' {\em
  \pasp}~{\bf 112},  1479--1486 (Nov. 2000).

\bibitem{Roddier}
{Roddier}, F. and {Roddier}, C., ``{Stellar Coronograph with Phase Mask},''
  {\em \pasp}~{\bf 109},  815--820 (July 1997).

\bibitem{Ruane_2018}
{Ruane}, G., {Riggs}, A., {Mazoyer}, J., {Por}, E.~H., {N'Diaye}, M., {Huby},
  E., {Baudoz}, P., {Galicher}, R., {Douglas}, E., {Knight}, J., {Carlomagno},
  B., {Fogarty}, K., {Pueyo}, L., {Zimmerman}, N., {Absil}, O., {Beaulieu}, M.,
  {Cady}, E., {Carlotti}, A., {Doelman}, D., {Guyon}, O., {Haffert}, S.,
  {Jewell}, J., {Jovanovic}, N., {Keller}, C., {Kenworthy}, M.~A., {Kuhn}, J.,
  {Miller}, K., {Sirbu}, D., {Snik}, F., {Wallace}, J.~K., {Wilby}, M., and
  {Ygouf}, M., ``{Review of high-contrast imaging systems for current and
  future ground- and space-based telescopes I: coronagraph design methods and
  optical performance metrics},'' in [{\em Space Telescopes and Instrumentation
  2018: Optical, Infrared, and Millimeter Wave}{\nolinebreak\hspace{0.1em}]},
  {Lystrup}, M., {MacEwen}, H.~A., {Fazio}, G.~G., {Batalha}, N., {Siegler},
  N., and {Tong}, E.~C., eds., {\em Society of Photo-Optical Instrumentation
  Engineers (SPIE) Conference Series} {\bf 10698},  106982S (Aug. 2018).

\bibitem{SLM}
Zhang, Z., You, Z., and Chu, D., ``Fundamentals of phase-only liquid crystal on
  silicon (lcos) devices,'' {\em Light: Science and Applications}~{\bf 3} (10
  2014).

\bibitem{Kuhn_2016}
{K{\"u}hn}, J. and {Patapis}, P., ``{Digital adaptive coronagraphy using SLMs:
  promising prospects of a novel approach, including high-contrast imaging of
  multiple stars systems},'' in [{\em Advances in Optical and Mechanical
  Technologies for Telescopes and Instrumentation
  II}{\nolinebreak\hspace{0.1em}]},  {Navarro}, R. and {Burge}, J.~H., eds.,
  {\em Society of Photo-Optical Instrumentation Engineers (SPIE) Conference
  Series} {\bf 9912},  99122M (July 2016).

\bibitem{Kuhn_2022}
K{\"u}hn, J.~G., Jolissaint, L., Bouxin, A., and Patapis, P., ``{SLM-based
  active focal-plane coronagraphy: status and future on-sky prospects},'' in
  [{\em Advances in Optical and Mechanical Technologies for Telescopes and
  Instrumentation IV}{\nolinebreak\hspace{0.1em}]},  Navarro, R. and Geyl, R.,
  eds.,  {\bf 11451},  114511S, International Society for Optics and Photonics,
  SPIE (2021).

\bibitem{dual_zone}
{Soummer}, R., {Dohlen}, K., and {Aime}, C., ``{Achromatic dual-zone phase mask
  stellar coronagraph},'' {\em \aap}~{\bf 403},  369--381 (May 2003).

\bibitem{wrap_vortex}
{Galicher}, R., {Huby}, E., {Baudoz}, P., and {Dupuis}, O., ``{A family of
  phase masks for broadband coronagraphy example of the wrapped vortex phase
  mask theory and laboratory demonstration},'' {\em \aap}~{\bf 635},  A11 (Mar.
  2020).

\end{thebibliography}
\bibliographystyle{spiebib} 

\end{document}